\documentstyle[twoside,fleqn,espcrc2,epsf]{article}

\title{Discovery of SAX J1753.5-2349 and SAX J1806.5-2215: two
	X-ray bursters without detectable steady emission}
\author{J.J.M.~in~'t~Zand\address{Space Research Organization
	Netherlands, Sorbonnelaan 2, 3584 CA Utrecht, 
	the Netherlands}, J. Heise$,^{\rm a}$
	J.M. Muller$^{\rm a,}$\address{BeppoSAX Science Data Center, 
	Nuova Telespazio, Via Corcolle 19, 00131 Roma, Italy},
	A. Bazzano$^{\rm c}$, M. Cocchi,$^{\rm c}$ L. Natalucci$^{\rm c}$, 
	P. Ubertini\address{
	Istituto di Astrofisica Spaziale (CNR), 
	Via del Fosso del Cavaliere, 00133 Roma, Italy}}
\begin{document}

\begin{abstract}
We report the discovery with BeppoSAX-WFC of two new X-ray sources that were 
only seen during bursts: SAX~J1753.5-2349 and SAX~J1806.5-2215. For both sources,
no steady emission was detected above an upper limit of 5~mCrab 
(2 to 8 keV) for 3~10$^5$~s around the burst events. The single burst detected 
from SAX~J1753.5-2349 shows spectral softening and a black body
color temperature of 2.0~keV. Following the analogy with bursts in other 
sources the burst very likely originates in a thermonuclear flash on
a neutron star. The first of two burst detected from SAX~J1806.5-2215 
does not show spectral softening and cannot be confirmed as a thermonuclear 
flash.
\end{abstract}

\maketitle

\section{Introduction}
The Wide Field Camera instrument (WFC, Jager et al. 1997) on board the
BeppoSAX satellite (e.g., Boella et al. 1997) has the largest field of view
(FOV)
of any astrophysical X-ray imaging device flown so far.
This implies an exceptional capability in finding short unexpected
transient sources with durations between seconds and
one hour, such as gamma and X-ray bursts. Only instruments that 
cover relevant portions of the sky with an appropriate
time coverage can be efficient in finding such events. The relevant 
sky portion for gamma-ray bursts is the whole sky, for
X-ray bursts this is the galactic bulge. According to a recent count
of X-ray bursters before the launch of BeppoSAX (Van Paradijs 1995), about 80\%
of all known X-ray bursters are within 20 degrees from the direction to
the galactic center. WFC can cover all of this region with a single 
pointing and it is
obvious it can contribute a lot to the knowledge of sources of X-ray bursts.

The galactic bulge region is regularly monitored with WFC as part of an
ongoing program. So far, there have been campaigns in the spring of 1997
and the falls of 1996 and 1997. In the first operational year of BeppoSAX 
over 10$^6$~s has been accumulated on the region and has turned up a number
of new X-ray bursters (In~'t~Zand et al., 1998a, In~'t~Zand et al. 1998b, 
Heise et al. 1997, Cocchi et al. 1997a).
Four of these were first detected as X-ray sources.
This paper presents two of those. They are set apart
by the fact that they were only detected during bursts,
no steady emission was detected. A general review of the 
bulge observations with WFC may be found in Heise~et~al.~1997.

The WFC instrument consists of two identical coded aperture cameras.
Each has a FOV of 40 by 40 square degrees and covers 3.7\%
of the sky. The angular resolution is 5' full width at half maximum.
The cameras are pointed in opposite directions. Because 
of the low-earth orbit of BeppoSAX the FOV of either camera is at any time 
usually blocked by the earth. The other X-ray instruments, the so
called narrow field instruments, view the sky perpendicular to the WFC.
Most of the galactic bulge observations with WFC are performed with
WFC as prime instrument and the center of the FOV of either camera is 
pointed as close as possible to the galactic center. The very bright and, 
therefore, disturbing source Sco~X-1 is usually kept outside the FOV.

\begin{figure}[t]
  \begin{center}
    \leavevmode
\epsfxsize=7cm
\epsfbox{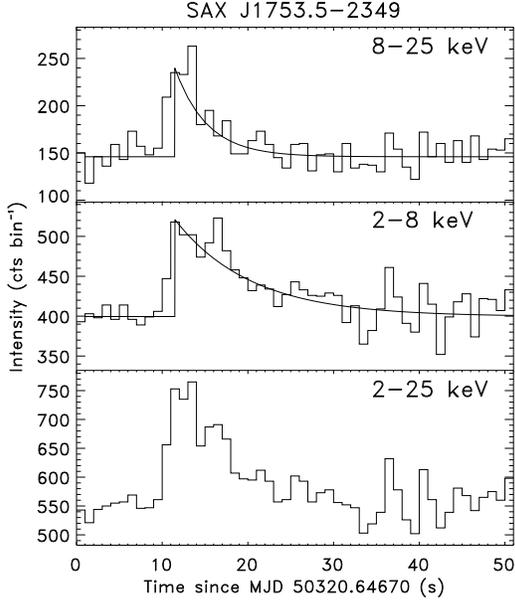}
  \caption{
Time history of SAX~J1753.5-2349 in two bandpasses and in the total
bandpass. The photons have been counted in the portion of
the detector illuminated by the source. No background subtraction was 
performed. The bin time is 1~s.\label{fig1753lc}
}
  \end{center}
\end{figure}

\begin{figure}[t]
  \begin{center}
    \leavevmode
\epsfxsize=7cm
\epsfbox{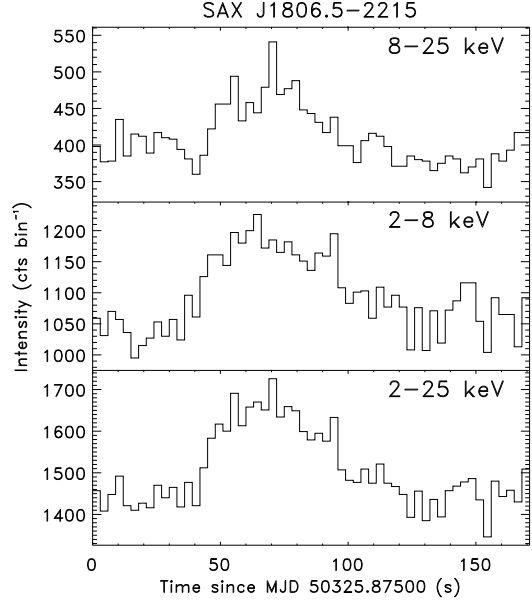}
  \caption{
Time history of SAX~J1806.5-2213 in two bandpasses and in the total
bandpass. See caption to Figure~\ref{fig1753lc} for explanation. The
bin time is 3~s.\label{fig1806lc}
}
  \end{center}
\end{figure}

\begin{figure}[t]
  \begin{center}
    \leavevmode
\epsfxsize=7cm
\epsfbox{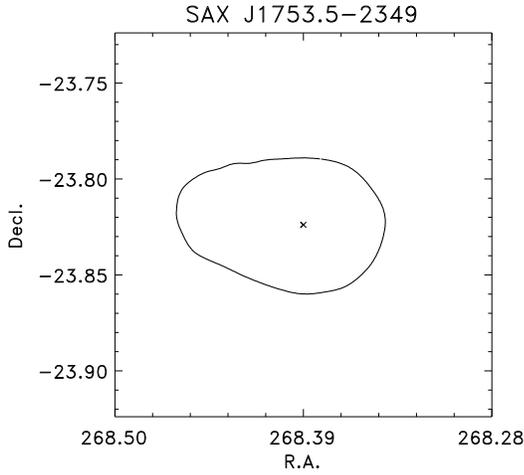}
  \caption{
Error region of SAX~J1753.5-2349 for a confidence level of 99\%. The cross
indicates the best fit position
\label{fig1753err}
}
  \end{center}
\end{figure}

\begin{figure}[t]
  \begin{center}
    \leavevmode
\epsfxsize=7cm
\epsfbox{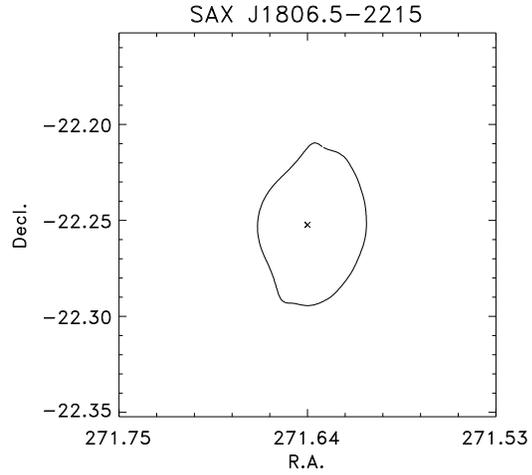}
  \caption{
Error region of SAX~J1806.5-2215 for a confidence level of 99\%, based
on the first burst. The cross indicates the best fit position
\label{fig1806err}
}
  \end{center}
\end{figure}

\section{SAX~J1753.5-2349}

SAX~J1753.5-2349 was detected during a single burst on August 24, 1996. The
time profile of this burst is shown in Figure~\ref{fig1753lc} in two bandpasses
and the error region is depicted in Figure~\ref{fig1753err}. The best fit 
position is at R.A.~=~17$^{\rm h}$53$^{\rm m}$34$^{\rm s}$, 
Decl.~=~$-23^{\rm o}$49.4' (J2000.0). No 
identification could be made with any other known object in any wavelength band 
within the error box using the Simbad database in December 1997. The peak 
intensity of the burst is 0.9~Crab units in 2 to 8 keV. 

There is strong proof of spectral softening. We fitted an exponential
function to the decaying part of the burst in 2 to 8 and 8 to 25 keV,
see Figure~\ref{fig1753lc}, and find 1/e decay times of 8.9$\pm$1.9~s
and 3.8$\pm0.7$~s respectively.

The spectral model could not be constrained. However, if one assumes a black 
body spectrum, like is common among X-ray burst spectra,
the color temperature is $2.0\pm0.2$~keV.

Not correcting for color nor gravitational redshift and assuming isotropic
radiation, the black body emission region can be characterized by 
a sphere with a radius of $11\pm2$~km/$d_{\rm 10 kpc}$ where $d_{\rm 10 kpc}$ 
is the distance to earth in units of 10~kpc.

The spectral softening and the black body color temperature, together
with the time and size scale identifies SAX~J1753.5-2349 with great
certainty as a low-mass X-ray binary with a neutron star as compact
object (see Lewin, Van Paradijs \& Taam, 1995, for these diagnostics).

\section{SAX~J1806.5-2215}

SAX~J1806.5-2213 was detected during two bursts on August 30, 1996, and
March 30, 1997. We here discuss the first burst which enables easier
analysis than the second one. The second burst is presented by Cocchi et al.
(1997a). The time profile of the first burst is presented in 
Figure~\ref{fig1806lc} and the error region in Figure~\ref{fig1806err}.
The burst is relatively long, about one minute, and has a peak intensity of
1.9 Crab in 2 to 8 keV which is also relatively bright.
The best fit position of the burst is 
R.A.~=~18$^{\rm h}$06$^{\rm m}$34$^{\rm s}$, 
Decl.~=~$-22^{\rm o}$15.1' (J2000.0). The error box does not contain
a known object in any wavelength band according to the Simbad database
in December 1997. 

The time profile in Figure~\ref{fig1806lc} shows that there is no noticeable 
spectral softening 
in the burst. This sets it apart from typical X-ray bursts due to thermonuclear
flashes and hampers easy characterization of this X-ray source. Nevertheless, 
if we fit a black body model to the observed spectrum, we find a color
temperature common for such X-ray bursts: $1.7\pm0.2$~keV. The
radius of the isotropically emitting sphere would be 14$\pm$3~km/$d_{\rm 10 kpc}$.

\section{Steady emission}

No steady emission was detected from both sources. For all observations of
the month August 1996 with a total exposure time of $3~10^5$~s the upper 
limit is 5~mCrab in 2 to 8 keV for both sources. SAX~J1753.5-2349 and
SAX~J1806.5-2215 are the only sources in
the complete WFC data set of over 10$^6$~s exposure time on the galactic 
bulge that have been seen only during bursts brighter than $\sim$0.3~Crab.
Probably it is coincidence that no steady emission was detected. WFC
is not an all-sky monitor and, therefore, has a far from complete
coverage of the two sources. They might have exhibited
steady emission above the WFC detection limit but just
before WFC observed the bursts. WFC observations of XTE~J1709-267 
(Cocchi et al. 1997b) exemplify this: X-ray bursts were seen from this 
object even when the steady emission dropped below the detection limit 
of WFC. If bright steady emission from SAX J1753.5-2349 or SAX J1806.5-2215 
would have existed before the WFC observations it could not have been 
above roughly 0.1~Crab units, though, because then it would very probably 
have been detected with the all-sky monitor ASM onboard the Rossi X-ray 
Timing Experiment (Levine et al. 1996) which is not the case.

\vspace{0.5cm}\noindent
{\it Acknowledgements.}
We thank A. Klumper, M. Savenije, J. Schuurmans and G. Wiersma at SRON for
software support during the analysis, and the staff of the
BeppoSAX {\em Satellite Operation Center} and {\em Science Data Center}
for the help in carrying out and processing
the Galactic Center observations with the WFC. The BeppoSAX satellite is
a joint Italian and Dutch program. A.B., M.C., L.N. and P.U. thank Agenzia
Spaziale Nazionale ASI for support. This research has made use of the Simbad
database, operated at CDS, Strasbourg, France.


\end{document}